\documentclass[prl,preprint,showpacs]{revtex4}   
\usepackage{amsmath,amsfonts}
\usepackage{epsfig}
\usepackage{graphicx}

\newcommand{\bea}{\begin{eqnarray}}
\newcommand{\beq}{\begin{equation}}
\newcommand{\eea}{\end{eqnarray}}
\newcommand{\eeq}{\end{equation}}
\def\diadots{\mathinner{\mkern1mu\raise8pt\vbox{\kern7pt\hbox{.}}\mkern2mu
    \raise2pt\hbox{.}\mkern2mu\raise-4pt\hbox{.}\mkern1mu}}

\begin{document}
\title
{An effective interaction at the Planck mass and the Planck length} 
\author {WD Heiss} 
\affiliation {Institute of Theoretical Physics, University Stellenbosch\\
National Institute for Theoretical Physics, South Africa }
\begin{abstract}
A simple model is suggested by assuming an effective interaction at the crossing point of the two curves,
the Schwarzschild radius and the Compton wave length, that is at the Planck length and Planck mass.
It is argued that there would be a physical effect that may be measurable at, say, nucleon mass and size,
while the Schwarzschild radius would remain unaffected for macroscopic lengths and masses.
\end{abstract}
\pacs{04.60.Bc, 04.80.-y}
\maketitle

It is well known that the Planck mass and the Planck length are determined as the
intersection point of the two curves: the Schwarzschild radius and the Compton wave length
when the respective lengths are plotted {\it versus} the mass. It is also known that
the Planck length is too minuscule to be amenable to observation while the Planck mass
is macroscopic. For the following we prefer to view these two curves with the axes
interchanged, i.e. we view the two curves with the mass plotted as a function of the
length. The character of the curves does not change: the straight line (Scharzschild radius)
remains a straight line and the hyperbola (Compton wave length) remains a hyperbola.
Note that in this redefined plot the observable physics is at the very far right
from the intersection point, in fact, the Planck length being about $10^{-35}[m]$ is 20 orders
of magnitude smaller than the size of a nucleon being about $10^{-15}[m]$.

The masses, now given at the ordinate, are a form of energy that can interact
(we may think of multiplying the mass by $c^2$).
When we assume an effective interaction at the intersection point, the crossing point
will become a level repulsion \cite{hs}.
Our interest here is exclusively focused upon the physical consequences
of such an interaction rather than upon the precise nature giving rise to such an effective
interaction. Also, whether or not the two levels cross or repel is at this stage beyond observation
even though it might be of fundamental interest. Somewhat related in spirit, but using a different
starting point, is a discussion about
'The Black Hole Uncertainty Principle Correspondence' as presented in \cite{carr}. 

The two curves, the Schwarzschild radius and the Compton wave length, are presented in the diagonal of the matrix
\beq
H=\begin{pmatrix} x/a & m_{\rm P}f \\ m_{\rm P}f & b/x \\
\end{pmatrix}
\label{mat}
\eeq
where $a=G/c^2$ and $b=\hbar/c$. The coupling term has been chosen to be a (small) multiple of the Planck
mass $m_{\rm P}=\sqrt{b/a}$ as the coupling term must also have the dimension of a mass. The dimensionless constant 
$f$ with $0 \le f \ll 1$  denotes the strength of the coupling. We are aware that
an effective interaction as used here will probably involve higher order terms of the Planck mass but this would require
more parameters which we do not consider in this simple treatment. In order to ensure that the gravitational mass
is always largest we consider the two curves only for $x\ge l_{\rm Planck}$ with $l_{\rm Planck}=\sqrt{ab}$ being the
Planck length.

The eigenvalues of $H$ are
\beq
m_{1,2}(x)=\frac{b}{2x}+\frac{x}{2a}(1 \pm \sqrt {1+\frac{a^2b^2}{x^4}-\frac{2ab}{x^2}+\frac{4abf^2}{x^2} }\,).
\label{eig} 
\eeq
As noted above we now focus our attention upon large values of $x$, that is we expand (\ref{eig}) in powers of $x$.
The leading term of the first eigenvalue is $x/a$, it is independent of the coupling. The next order $bf^2/x$
is minuscule as such and even more so for large values of $x$. In other words, the Schwarzschild radius remains
essentially unaffected by the coupling.

\begin{figure}
\includegraphics[height=0.30\textheight]{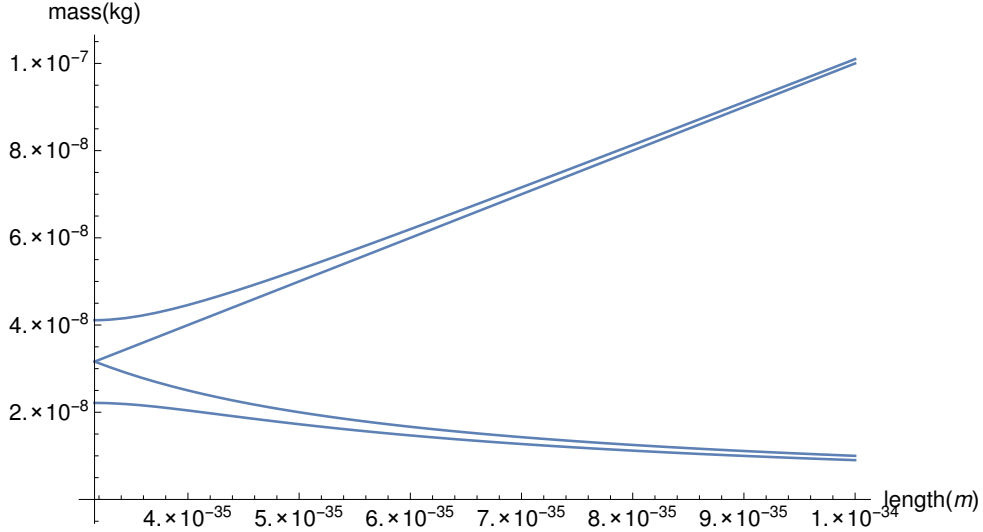}
\caption{The two sets of curves, the one for $f=0$ where the straight line and the hyperbola are joining at the left 
at $l_{\rm Planck}$, and the corresponding
curves with $f=0.3$ as chosen here for illustration. The rough orders of magnitudes as given in the text for $a$ and $b$ are used here. }
\end{figure}

However, the second eigenvalue is affected by the coupling. The expansion yields for the leading term
\beq
m_2(x)=\frac{b}{x}(1-f^2) +O(1/x^3).
\label{exp}
\eeq
The higher orders have the form $b\, l_{\rm Planck}^{2n}/x^{2n+1}P_{n+1}(f^2), n=1,2,\ldots $ where $P_n$ is a polynomial
that vanishes at $f=0$ (and at $f=1$). These terms again are minuscule. Note that
the higher orders depend explicitly on $G$ while (\ref{exp}) appears to be independent of $G$, but this is only
apparent, since $f$ is a fraction of the Planck mass which does depend on $G$.

Observable physics happens some twenty orders of magnitude further to the right of the display in Fig.1.
Inserting simply orders of magnitudes for $a=G/c^2\approx 10^{-27}$ and $b=\hbar/c\approx 10^{-42}$ (in [m,kg,s]) we find for $x=10^{-15}$ 
and $f=0$ the mass $10^{-27}$ which becomes smaller by the factor $(1-f^2)$ when $f$ is switched on. 
Moving towards smaller values of $x$ by the same factor the original mass value is retrieved.

In other words, the interpretation of (\ref{exp}) is that the product of the Compton wave length (denoted here by $x$) of, say,
a proton and its mass is diminished by the factor $(1-f^2)$. Fixing the mass as measured by experiment the
associated Compton wave length will appear smaller by the factor $(1-f^2)$. 

It is not obvious from this consideration whether the interaction at the Planck length as assumed in this paper
diminishes the mass or the Compton wave length or both of, say, a nucleon as only the product is affected. 
If the factor $f$ is not too small, the deviation should be measurable.
In addition, as the consideration shows, the mere assumption of an interaction at the Planck length will have a physical 
effect in the realms of quantum mechanics whereas the effect would be virtually zero for the gravitational attraction
between masses.  

\section*{Acknowledgment}
The author gratefully acknowledges helpful and pertinent comments by Professor Robert de Mello-Koch.

\end{document}